\documentclass[twocolumn,dvips,showpacs,showkeys,aps,pra]{revtex4} % twocolumn
\usepackage{graphicx}% Include figure files
\usepackage{dcolumn}% Align table columns on decimal point
\usepackage{bm}% bold math
\input{epsf}
\usepackage{amsmath,amsfonts,amsthm,dcolumn,pstricks,pst-plot}
\usepackage{epsfig,subfigure}
\usepackage[mathscr]{eucal}
\usepackage{mathptmx}
\newcommand{\be}{\begin{equation}}
\newcommand{\ee}{\end{equation}}
\begin{document}

\title{Exact results for the Wigner transform phase space densities of a two--dimensional\\
harmonically confined charged quantum gas subjected to a
magnetic field}
\author{K. Bencheikh}
\email{bencheikh.kml@gmail.com}
\affiliation{D\'{e}partement de Physique. Laboratoire de physique quantique et syst\`{e}%
mes dynamiques. Universit\'{e} de S\'{e}tif, Setif 19000, Algeria}
\author{L.M. Nieto}
\email{luismi@metodos.fam.cie.uva.es}
\affiliation{Departamento de F\'{\i}sica Te\'{o}rica, At\'{o}mica y \'{O}ptica,
Universidad de Valladolid, 47071 Valladolid, Spain }
\date{\today}

\begin{abstract}
Closed form analytical expressions are obtained for
the Wigner transform of the Bloch density matrix and for the Wigner phase
space density of a two dimensional harmonically trapped charged quantum gas
in a uniform magnetic field of arbitrary strength, at zero and nonzero
temperatures. An exact analytic expression is also obtained for the
autocorrelation function. The strong magnetic field case, where only few
Landau levels are occupied, is also examined, and useful approximate
expressions for the spatial and momentum densities are given.
\end{abstract}


\maketitle

%%%%%%%%%%%%%%%%%%%%%%%%%%%%%%%%%%%%%%%%%%%%%%%

%\newpage

%\baselineskip=15pt

\section{Introduction}

Considerable interest has been shown in the study of the properties of the
so-called low dimensional systems. The advances in nanotechnology allows
nowdays the realisation of quasi-two dimensional systems like quantum dots 
\cite{[1],[2]}. In a different context, the experimental achievement
of trapped ultra-cold atom gases allows to study quantum mechanical effects
of quantum statistics in such gases \cite{[3]}. The above mentionned
physical systems have originated a great volum of theoretical work in order
to understand such fascinating world in a reduced physical space \cite{[4]}.
In this context, using the canonical Bloch density matrix as a tool, exact
analytical expressions have been obtained for the particle and the kinetic
energy densities in spatial coordinates at zero and nonzero temperatures 
\cite{[5]}. Very recently, this method has been generalized to take into
account the effect of a uniform perpendicular magnetic field on a confined
charged two dimensional quantum gas \cite{[6]}. In the present work, we are
interested in obtaining exact analytical expressions for the Wigner
transforms of both, the canonical Bloch density matrix and the first-order
density matrix. Our interest in the Wigner tranform is based on the fact
that it provides a useful tool to study various properties of many-body
systems \cite{[7]}. Besides the fact that it allows a reformulation of
quantum mechanics in terms of classical concepts, and it is also used to
generate semi classical approximations \cite{[8],[9]}, the Wigner
transform may help to gain a better understanding of the properties of the
system. It is also very interesting because with the recent progress of the
experimental techniques, phase space densities can be nowdays measured for
certain quantum systems \cite{[10]}.

The canonical Bloch density matrix is defined as $C({\bm r},%
{\bm r}' ,\beta )=\sum_{j}\varphi _{j}({\bm r%
})\varphi_{j}^{\ast }({\bm r}' )\exp (-\beta \epsilon _{j})$,
where $\varphi _{j}({\bm r})$ and $\epsilon _{j}$ are
eigenfunctions and eigenvalues of a one particle Hamiltonian $H$ associated
to the system. Here, $\beta $ is to be interpreted as a mathematical
variable, which in general, is taken to be complex, and not necesarily the
inverse temperature.
The Bloch density matrix $\ $is of particular interest since its knowledge
enables the first-order density matrix $\rho ({\bm r},%
{\bm r}')$ to be found, through the inverse Laplace transform 
\cite{[11]}. In fact at $T=0$, the first-order density matrix, for a given
Fermi energy $\lambda $, is given by%
\begin{equation}
\rho ({\bm r},{\bm r}')=\frac{1}{2\pi i}%
\int_{c-i\infty }^{c+i\infty }d\beta\ \frac{C({\bm r},{\bm r}' , \beta )}{\beta} \ e^{\beta\lambda } .
%\tag{1}
\end{equation}
The system we are going to study is a harmonically confined charged atom gas
in a two dimensional $x$--$y$ plane subjected to a perpendicular homogeneous
magnetic field ${\bm B}=B\,{\bm k}$,\ taken along the $z$
axis. The one particle Hamiltonian is then given by 
\begin{equation}
H=\frac{1}{2m^{\ast }}\left( \frac{\hbar }{i}{\bm \nabla }+\frac{e%
}{c}{\bm A}\right) ^{2}+\frac{1}{2}m^{\ast }\omega _{0}^{2}%
{\bm r}^{2}  %\tag{2}
\end{equation}%
with ${\bm r}^{2}=x^{2}+y^{2}$, ${\bm A}$ =$\frac{1}{2} ({\bm B}\times {\bm r})$ is the vector potential, $%
m^{\ast }$\ and $-e$ are respectively the effective mass and the charge of
the particle and $\omega _{0}$ is the oscillation frequency of the confining
potential. For the Hamiltonian under study, a closed analytical expression
was obtained long time ago for the corresponding Bloch density \cite{[12]}.
Here we rewrite it in the following useful form%
\begin{widetext}
\be
C({\bm r},{\bm r}' ,\beta ) =
\frac{m^{\ast}\Omega/2\pi \hbar  }{\sinh \beta \hbar \Omega}\exp \left\{ \frac{
-2m^{\ast }\Omega/\hbar }{ \sinh \beta \hbar \Omega}\left[ % \left( 
{\bm R}^{2} \sinh \frac{\beta \hbar \Omega _{-}}{2}\sinh \frac{\beta \hbar \Omega _{+}}{2} +
\frac{{\bm s}^{2}}{4} \cosh \frac{\beta \hbar \Omega _{-}}{2}\cosh \frac{\beta
\hbar \Omega _{+}}{2}  + i\frac{ ({\bm R}\times 
{\bm s})\cdot {\bm k}}{2} \sinh \beta \hbar \omega _{L} \right] \right\} 
%\tag{3}
\ee
\end{widetext}
where ${\bm R}=({\bm r}+{\bm r}')/2$ and 
${\bm s}={\bm r}-{\bm r}'$ are,
respectively, the center of mass and relative coordinates, and 
\begin{equation}
\omega _{L}=\frac{eB}{2m^{\ast }c},\
\Omega =\sqrt{\omega _{0}^{2}+\omega _{L}^{2}}, \
\Omega _{\pm }=\Omega \pm\omega _{L}. 
%\tag{4}
\end{equation}%
$\omega _{L}$ is the Larmor frequency and $\Omega _{\pm }$ are two
frequencies that correspond to excitations in the center of mass
motion--the so-called ``Kohn modes'' \cite%
{[13]}. Note that, the Hamiltonian in Eq.~(2) has the same partition
function, $Z=1/\left[ 4\sinh (\beta \hbar \Omega _{+}/2)\sinh (\beta \hbar
\Omega _{-}/2)\right] $, as that an anisotropic two dimensional harmonic
oscillator with frequencies $\Omega _{-}$ and $\Omega _{+}$.

The rest of the paper is organized as follows. In the next section, we
calculate the Wigner transform of the Bloch density, and alternative useful
analytical forms for such Wigner transfom are also derived. The Wigner phase
space density matrix is calculated at zero and nonzero temperatures in
section 3, showing some interesting plots of it. In section 4, we derive a closed analytical form, in terms of
Laguerre polynomials, for the so called autocorrelation function. The high
magnetic field strength case is examined in section 4. In the last section,
a summary and outlook are given.

\section{The Wigner Transform of the Bloch density matrix}

In the following we shall calculate the Wigner transform of the Bloch
density matrix given in Eq. (3). The Wigner transform of an arbitrary one particle operator $A$, defined by its matrix elements in spatial coordinates $A({\bm r}+%
\frac{{\bm s}}{2},{\bm r}-\frac{{\bm s}}{2}%
) $, is the following function $A_W$ of the phase space variables $%
{\bm r}$ and ${\bm p}$ \cite{[14]} 
\begin{equation}
A_{W}({\bm r},{\bm p})=\int_{\mathbb R^2} A({\bm r}+{\bm s}/2,{\bm r}-{\bm s}/2)e^{-i{\bm p}\cdot{\bm s}/\hbar }\ d{\bm s},  %\tag{5}
\end{equation}%
being its inverse transform 
\begin{equation}
A({\bm r}+{\bm s}/2,{\bm r}-{\bm s}/2)=\int_{\mathbb R^2} \frac{d{\bm p}}{(2\pi \hbar )^{2}}\ A_{W}({\bm r},{\bm p})e^{i{\bm p}\cdot {\bm s}/\hbar }\mathbf{.}  %\tag{6}
\end{equation}%
According to Eq. (6), the local part of the operator $A$ can be computed as 
\begin{equation}
A({\bm r},{\bm r})\equiv A({\bm r})=\int_{\mathbb R^2} \frac{d{\bm p}}{(2\pi \hbar )^{2}}A_{W}({\bm r},
{\bm p}) .  %\tag{7}
\end{equation}%
We can now calculate, by making use of Eq. (5),
the Wigner transform of the Bloch density matrix (3). Let us take the Wigner
transform of $C({\bm r}+\frac{{\bm s}}{2},%
{\bm r}-\frac{{\bm s}}{2},\beta )$ and call it $C_{W}(%
{\bm r},{\bm p},\beta )$, so that%
\begin{widetext}
\begin{eqnarray}
C_{W}({\bm r},{\bm p},\beta ) &=&\frac{m^{\ast }\Omega/2\pi \hbar
}{ \sinh \beta \hbar \Omega}\ \exp \left[ -\frac{2m^{\ast }\Omega 
}{\hbar \sinh (\beta \hbar \Omega )}\sinh \frac{\beta \hbar \Omega _{-}}{2}%
\sinh \frac{\beta \hbar \Omega _{+}}{2}\ {\bm r}^{2}\right] 
\notag \\
&&\qquad \int_{\mathbb R^2}  \exp \left[ -\frac{m^{\ast }\Omega }{%
2\hbar \sinh (\beta \hbar \Omega )}\left( \cosh \frac{\beta \hbar \Omega _{-}%
}{2}\cosh \frac{\beta \hbar \Omega _{+}}{2}\right) {\bm s}%
^{2}
-i\left( \frac{m^{\ast }\Omega \sinh \beta \hbar \omega _{L}}{\hbar \sinh
(\beta \hbar \Omega )}({\bm k}\times {\bm r})+\frac{%
{\bm p}}{\hbar }\right) \cdot {\bm s} \right] \ d{\bm s} .  %\tag{8}
\end{eqnarray}%
\end{widetext}
The above two dimensional integral can be easily evaluated by using the well known identity
\be
\int_{{\Bbb R}^2} d{\bm s}\ e^{-a{\bm s}^{2}-i{\bm b}\cdot {\bm s}}=\frac{\pi }{a}e^{-{\bm b}^{2}/(4a)} 
\ee
to obtain the result
\begin{equation}
C_{W}({\bm r},{\bm p},\beta )=\frac{e^{-f(\beta){\bm r}^{2}}\ e^{-g(\beta)\left( 
{\bm p}+u(\beta)({\bm k}\times {\bm r})\right) ^{2}}}{\cosh \frac{%
\beta \hbar \Omega _{+}}{2}\cosh \frac{\beta \hbar \Omega _{-}}{2}}%
 ,  %\tag{10}
\end{equation}%
where for notational simplicity we have introduced the following functions
of $\beta $
\begin{eqnarray}
 f(\beta)&=&\frac{2m^{\ast }\Omega \sinh \frac{\beta \hbar \Omega _{+}}{2}\sinh 
\frac{\beta \hbar \Omega _{-}}{2}}{\hbar \sinh (\beta \hbar \Omega )}, \nonumber \\
 g(\beta)&=& \frac{\sinh (\beta \hbar \Omega )}{2m^{\ast }\hbar \Omega \cosh \frac{\beta
\hbar \Omega _{+}}{2}\cosh \frac{\beta \hbar \Omega _{-}}{2}},\\
u(\beta)&=&\frac{m^{\ast }\Omega \sinh \beta \hbar \omega _{L}}{\sinh (\beta \hbar
\Omega )} .  \nonumber %\tag{11}
\end{eqnarray}%
It can be easily checked that when the magnetic field is absent, so that $%
\omega _{L}=0$ then $\Omega _{+}=\Omega _{-}=\omega _{0}$ and $\Omega
=\omega _{0}$, Eq. (10) yields to the correct Wigner transfom for a harmonic
oscillator in two dimensions, that is \cite{[8]}%
\begin{equation}
C_{W}^{B=0}({\bm r},{\bm p},\beta )=\frac{\exp\displaystyle \left[ -\frac{2 \tanh \frac{\beta \hbar \omega _{0}}{2}}{\hbar \omega _{0}}
\left(  \frac{{\bm p}^{2}}{2m^{\ast }}+\frac{m^{\ast }\omega _{0}^{2}}{2}
{\bm r}^{2} \right) \right]}{\cosh ^{2}
\frac{\beta \hbar \omega _{0}}{2}}  %\tag{12}
\end{equation}
For the case of an unconfined system subjected to a magnetic field, i.e. 
$\omega _{0}=0$, then $\Omega =\omega _{L}$, $\Omega _{-}=0$, $\Omega _{+}=2\omega _{L}$, and Eq. (10) reduces to 
\begin{equation}
C_{W}^{\omega _{0}=0}({\bm r},{\bm p},\beta )=\frac{\displaystyle\exp \left[ -\frac{\tanh \beta \hbar \omega
_{L}}{2m^{\ast } \hbar \omega _{L}} \left( {\bm p}+\frac{e%
}{c}{\bm A}\right) ^{2}  \right]}{%
\cosh \beta \hbar \omega _{L}} , %\tag{13}
\end{equation}%
which is the correct expression of the Wigner transform \cite{[15]}.

\subsection{Alternative forms for the Wigner transform of the Bloch density
matrix}

In the following, we present two alternative analytical forms of the
result in Eq. (10), which can be rewritten as 
\begin{equation}
C_{W}({\bm r},{\bm p},\beta )=\frac{1}{\cosh \frac{%
\beta \hbar \Omega _{+}}{2}\cosh \frac{\beta \hbar \Omega _{-}}{2}}\exp %
\left[ -G\left( {\bm r},{\bm p},\beta \right) \right] 
%\tag{14}
\end{equation}%
with 
\begin{equation}
G\left( {\bm r},{\bm p},\beta \right) =\left(
f+gu^{2}\right) {\bm r}^{2}+g{\bm p}^{2}+2guL_{z} 
%\tag{15}
\end{equation}%
and $L_{z}$ is the
component of the orbital angular momentum along the $z$ axis. We
show in Appendix A that $G\left( {\bm r},{\bm p}%
,\beta \right) $ takes the following simple form%
\begin{equation}
G\left( {\bm r},{\bm p},\beta \right) =\frac{H_0+\Omega L_{z}}{%
\hbar \Omega } \ \tanh \frac{\beta \hbar \Omega _{+}}{2}   + \frac{H_0-\Omega L_{z}}{\hbar \Omega }\ \tanh \frac{\beta \hbar \Omega _{-}}{2}. %\tag{16}
\end{equation}%
Substitution of this result into Eq. (15), leads to the factorized
analytical form 
%\begin{widetext}
\begin{eqnarray}
C_{W}({\bm r},{\bm p},\beta )&=&\frac{\exp \left[ -\frac{%
\tanh \frac{\beta \hbar \Omega _{+}}{2}}{\hbar \Omega }\left( H_0+\Omega L_{z}\right) \right] }{\cosh \frac{\beta \hbar
\Omega _{+}}{2}}\nonumber \\
&& \times  \frac{\exp \left[ -\frac{\tanh \frac{\beta \hbar
\Omega _{-}}{2}}{\hbar \Omega }\left( H_0-\Omega L_{z}\right) %
\right] }{\cosh \frac{\beta \hbar \Omega _{-}}{2}}  %\tag{17}
\end{eqnarray}%
%\end{widetext}
where
\begin{equation}
H_0= \frac{{\bm p}^{2}}{2m^{\ast }}+\frac{m^{\ast }\Omega^{2}}{2} {\bm r}^{2}.
\end{equation}%
Let us now obtain a third closed expression for Wigner transform of the
Bloch density. For that purpose, we use the following expansion in terms of Laguerre
polynomials \cite{[16]}%
\begin{equation}
\frac{\exp ( -x\tanh y) }{\cosh y} =2e^{-x}\sum_{n=0}^{\infty }(-1)^{n}L_{n}(2x)\exp \left[ -2 y (n+\frac{1}{2})\right]  %\tag{18}
\end{equation}%
for $x=\left( H_0\pm \Omega L_{z}\right) /\hbar \Omega $ and $y={\beta \hbar \Omega _{\pm }}/{2}$, Eq.~(17) becomes%
\begin{widetext}
\begin{equation}
C_{W}({\bm r},{\bm p},\beta )=4\exp \left[ -\frac{%
2H_{0}}{\hbar \Omega }\right] \sum_{n=0}^{\infty
}\sum_{m=0}^{\infty }(-1)^{n+m}L_{n}\left( \frac{2(H_{0}+\Omega
L_{z})}{\hbar \Omega }\right) L_{m}\left( \frac{2(H_{0}-\Omega L_{z})}{\hbar
\Omega }\right) \exp \left( -\beta E_{n,m}\right)  %\tag{19}
\end{equation}%
\end{widetext}
where
\begin{equation}
E_{n,m}=\hbar \Omega _{+}(n+1/2)+\hbar \Omega _{-}(m+1/2)  %\tag{21}
\end{equation}%
are the eingenvalues of the Hamiltonian (2). To our knowledge, the results given in Eqs. (10), (17) and (20) are new and seem not to have been reported before in the literature.

\section{Quantum Wigner phase space distribution at zero and nonzero temperatures}

\subsection{Phase space distribution at zero temperature}

Having established in the previous section various analytical forms for the
Wigner transform of the Bloch density, we shall now calculate analytically
the expression for the quantum Wigner phase space distribution or Wigner
transform density of the first-order density matrix $\rho ({\bm r}, {\bm r}')$. 
Let $\rho _{W}({\bm r},{\bm p})$ denote such density, defined as%
\begin{equation}
\rho _{W}({\bm r},{\bm p})=\int_{\mathbb R^2} \rho ({\bm r}+\frac{{\bm s}}{2},%
{\bm r}-\frac{{\bm s}}{2})e^{-i{\bm p}\cdot
{\bm s}/\hbar }\ d{\bm s},  %\tag{22}
\end{equation}%
The above distribution can also be obtained through the use of the Wigner
phase space version of (1), that is 
\begin{equation}
\rho _{W}({\bm r},{\bm p})=\frac{1}{2\pi i}%
\int_{c-i\infty }^{c+i\infty }d\beta\ \frac{C_{W}({\bm r},%
{\bm p},\beta )}{\beta }\ e^{\beta \lambda }  %\tag{23}
\end{equation}%
Inserting Eq. (20) into Eq. (23), and performing the inverse Laplace
transform \cite{[17]}%
\begin{equation}
\frac{1}{2\pi i}\int_{c-i\infty }^{c+i\infty }\frac{d\beta }{\beta }%
e^{\beta (\lambda -E_{n,m})}=\Theta (\lambda -E_{n,m}),  %\tag{24}
\end{equation}%
where $\Theta $ is the Heaviside step function, we find%
\begin{widetext}
\begin{equation}
\rho _{W}({\bm r},{\bm p})=4e^{-\frac{2H_{0}}{\hbar
\Omega }}\sum_{n=0}^{\infty }\sum_{m=0}^{\infty
}(-1)^{n+m}L_{n}\left( \frac{2(H_{0}+\Omega L_{z})}{\hbar \Omega }\right)
L_{m}\left( \frac{2(H_{0}-\Omega L_{z})}{\hbar \Omega }\right) \Theta
(\lambda -E_{n,m}) .  %\tag{25}
\end{equation}%
\end{widetext}
Due to the presence of the step function, the quantum numbers $n,m$ are restricted to $\hbar \Omega
_{+}(n+1/2)+\hbar \Omega _{-}(m+1/2)<\lambda $. The highest allowed value for $n$, $N_{+}$, is given by 
\begin{equation}
N_{+}=\text{Int}\left[ \frac{\lambda }{\hbar \Omega _{+}}-\frac{\Omega }{\Omega _{+}%
}\right] ,  %\tag{26}
\end{equation}%
where $\text{Int}(x)$ denotes the integer part of $x>0$. For a
given allowed value of $n$, the maximum allowed value  of $m$, $N_{-}$, is
\begin{equation}
N_{-}=\text{Int}\left[ \frac{\lambda }{\hbar \Omega _{-}}-\frac{\Omega _{+}}{\Omega
_{-}}n-\frac{\Omega }{\Omega _{-}}\right] .  %\tag{27}
\end{equation}%
Therefore, the density distribution in Eq. (25), can be rewritten as%
\begin{eqnarray}
\rho _{W}({\bm r},{\bm p})&=&4e^{-\frac{2H_{0}}{\hbar
\Omega }}\sum_{n=0}^{N_{+}}\sum_{m=0}^{N_{-}}(-1)^{n+m}L_{n}%
\left( \frac{2(H_{0}+\Omega L_{z})}{\hbar \Omega }\right) \nonumber \\
&& \qquad\qquad\qquad \times L_{m}\left( \frac{%
2(H_{0}-\Omega L_{z})}{\hbar \Omega }\right).  %\tag{28}
\end{eqnarray}
Notice that the above density depends not only on the moduli $|{\bm r}|$ and $|{\bm p}|$, but also on the relative angle  $\theta$ between ${\bm r}$ and ${\bm p}$, i.e., $\rho _{W}({\bm r},{\bm p})=\rho _{W}({r},{p},\theta)$. For $N=20$ particles, Fermi energy $\lambda=6.35\hbar\omega_L$  and $\omega_0/\omega_L=1$, we display in Fig.~1 (a) to (d) this Wigner phase space density for $\theta= 0,\pi/6,\pi/3,\pi/2$, respectively. 

%\begin{figure}[ht]%
%\centering
%\subfigure[]{%
%\includegraphics[width=5cm]%
%{wigner0.eps}
%} \qquad \qquad \qquad
%\subfigure[]{%
%\includegraphics[width=5cm]
%{wigner1.eps}
%}
%\\
%\subfigure[]{%
%\includegraphics[width=5cm]%
%{wigner2.eps}
%}  \qquad \qquad \qquad
%\subfigure[]{%
%\includegraphics[width=5cm]%
%{wigner3.eps}
\begin{figure}[ht]%
\centering
\subfigure[]{%
\includegraphics[width=5cm]%
{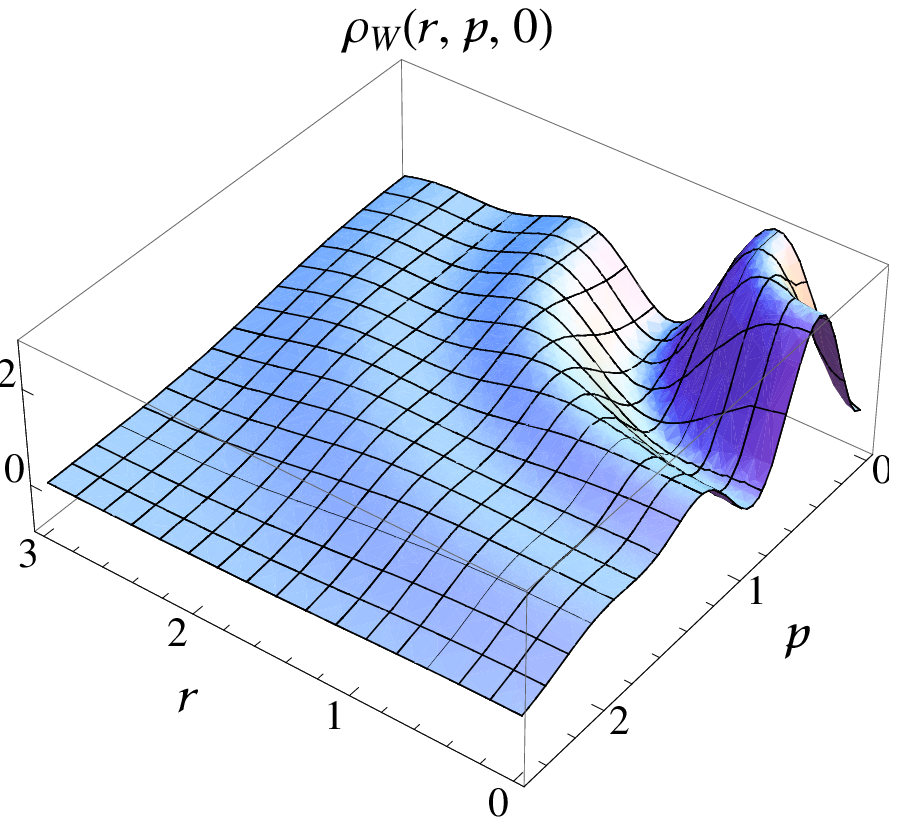}
} \qquad \qquad \qquad
\subfigure[]{%
\includegraphics[width=5cm]
{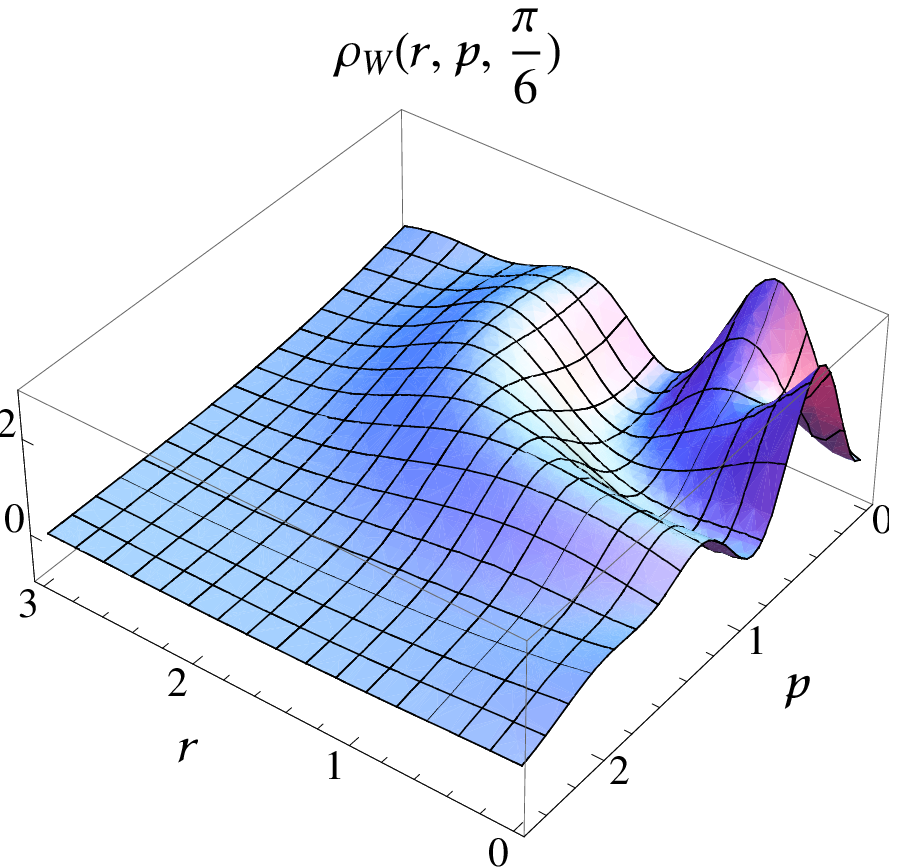}
}
\\
\subfigure[]{%
\includegraphics[width=5cm]%
{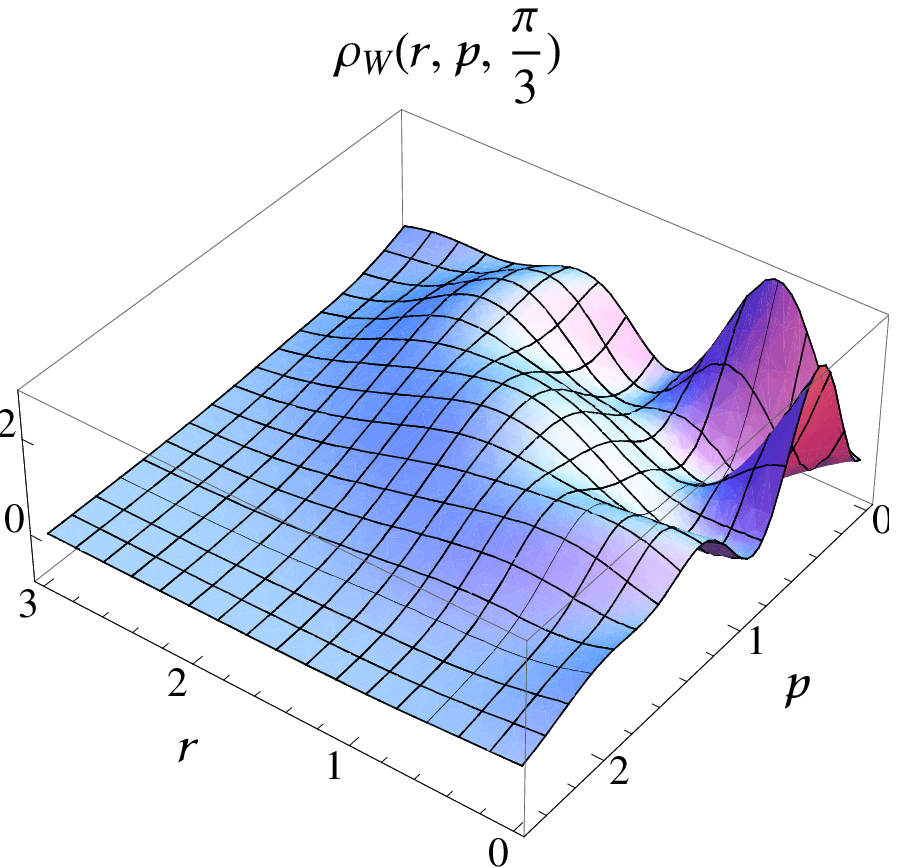}
}  \qquad \qquad \qquad
\subfigure[]{%
\includegraphics[width=5cm]%
{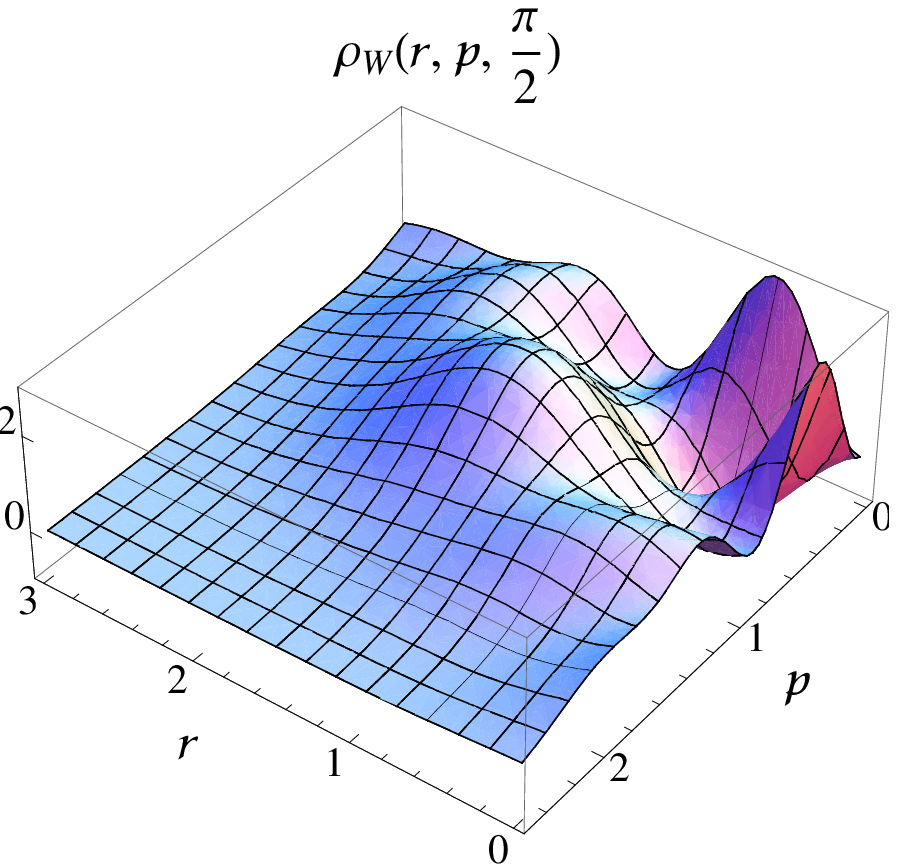}
}
\caption{Figures (a) to (d) correspond to plots of the Wigner phase space density $\rho _{W}({\bm r},{\bm p})=\rho _{W}({r},{p},\theta)$ for $N=20$ particles at $\theta= 0,\pi/6,\pi/3,\pi/2$, respectively. We have chosen parameters $\omega_0/\omega_L=1$ with Fermi energy $\lambda=6.35\hbar\omega_L$. Lengths are plotted in units of the magnetic length $l=\sqrt{\hbar c/eB}$ for $r$ and in units of $l^{-1}$ for the momentum $p$.}
\end{figure}

In the absence of a magnetic field, the system becomes of a pure harmonic
oscillator with frequency $\omega _{0}$ and thus all the $\Omega$ are all equal to $\omega _{0}$, we show in Appendix B, that the above
density has indeed the correct limit, given by \cite{[18]} 
\begin{equation}
\rho _{W}^{B=0}({\bm r},{\bm p})=4e^{-\frac{2H_{0}}{%
\hbar \omega _{0}}}\sum_{p=0}^{M}(-1)^{p}L_{p}^{1}\left( \frac{4H_{0}%
}{\hbar \omega _{0}}\right) .  %\tag{29}
\end{equation}%
where $L_{n}^{1}$ is the generalized Laguerre polynomial of order one and
the quantum number $M$ is related to the Fermi energy by $\lambda =\hbar
\omega _{0}(M+1)$.

In what follows, we shall deduce from Eq. (25) a result which will greatly
simplify us the treatment when we will deal, in the next subsection, with
the finite temperature case. Let $\phi _{n,m}({\bm r})$ denote
the eigenfunction of the Hamiltonian (2) with eigenvalues $E_{n,m}$
given in Eq. (21). In terms of the single particle wavefunctions, the
first-order density matrix is 
\begin{equation}
\rho ({\bm r}+\frac{\bm s}2,{\bm r}-%
\frac{\bm s}2)=\sum_{m,n=0}^{\infty } \phi _{n,m}({\bm r}+\frac{\bm s}2)\phi _{n,m}^{\ast }({\bm r}-\frac{\bm s}2)\Theta (\lambda -E_{n,m}),
%\tag{30}
\end{equation}%
where $\lambda $ is the Fermi energy. Taking the Wigner transform of Eq.
(30), we get 
\begin{equation}
\rho _{W}({\bm r},{\bm p})=\sum_{m,n=0}^{\infty
} {\mathcal W}\left[ \phi _{n,m}({\bm r}+\frac{\bm s}2) \phi _{n,m}^{\ast }({\bm r}-\frac{\bm s}2)\right] \Theta (\lambda -E_{n,m}).   %\tag{31}
\end{equation}%
Here the symbol ${\mathcal W}$ stands for Wigner transform. Comparing this result with
Eq. (25), we deduce that 
\begin{eqnarray}
\!\!\! &\!\!\! & {\mathcal W}\left[ \phi _{n,m}({\bm r}+\frac{\bm s}2)\phi
_{n,m}^{\ast }({\bm r}-\frac{\bm s}2)\right] = 4e^{-\frac{%
2H_{0}}{\hbar \Omega }}(-1)^{n+m} \nonumber \\
&& \qquad\qquad \times L_{n}\left( \frac{2(H_{0}+\Omega L_{z})}{%
\hbar \Omega }\right) L_{m}\left( \frac{2(H_{0}-\Omega L_{z})}{\hbar \Omega }%
\right) .  %\tag{32}
\end{eqnarray}%
Thus, we have found the Wigner transform of the product $\phi _{n,m}(%
{\bm r}+{\bm s}/2)\phi _{n,m}^{\ast }({\bm r%
}-{\bm s}/2)$ without the explicit use of the single particle
wavefunctions. As stated before this result will immediately be used in the
following subsection.

\subsection{Wigner phase space distribution at nonzero temperatures.}

Here, we shall generalize the result obtained in (25), valid for 
$T=0$, to nonzero temperatures. We start with the definition of the
first-order density matrix $\rho ({\bm r}+{\bm s}/2,%
{\bm r}-{\bm s}/2,T)$ at temperature $T$ in terms of
the normalized single particle wave functions $\phi _{n,m}$ , which reads for
Fermions 
\begin{equation}
\rho ^{F}({\bm r}+\frac{\bm s}2,{\bm r}-\frac{\bm s}2,T)=\sum_{n=0}^{\infty }\sum_{m=0}^{\infty}
\frac{\phi _{n,m}({\bm r}+\frac{\bm s}2)\phi _{n,m}^{\ast }(%
{\bm r}-\frac{\bm s}2)}{\exp \left( \frac{%
E_{n,m}-\mu }{k_{B}T}\right) +1},  %\tag{33}
\end{equation}%
where $\left[ \exp \left( \frac{%
E_{n,m}-\mu }{k_{B}T}\right) +1\right] ^{-1}$is the Fermi distribution
function for the level energy $E_{n.m}$, $k_{B}$ is Boltzmann's constant and $\mu $ the chemical potential.
Taking the Wigner transform of both sides in Eq. (33) and using obvious
notations, we get
\begin{equation}
\rho _{W}^{F}({\bm r},{\bm p},T)=\sum_{n=0}^{+%
\infty }\sum_{m=0}^{\infty } \frac{{\mathcal W}\left[ \phi _{n,m}({\bm r}+%
{\bm s}/2)\phi _{n,m}^{\ast }({\bm r}-{\bm s%
}/2)\right]}{\exp \left( \frac{E_{n,m}-\mu }{k_{B}T}\right) +1}, 
%\tag{34}
\end{equation}%
where we have used the fact that, the Fermi distribution is not affected by
the Wigner transformation. Substituting Eq. (32) into Eq. (34), one arrives
at%
\begin{widetext}
\begin{equation}
\rho _{W}^{F}({\bm r},{\bm p},T)=4e^{-\frac{2H_{0}}{%
\hbar \Omega }}\sum_{n=0}^{\infty }\sum_{m=0}^{\infty
}(-1)^{n+m}L_{n}\left( \frac{2(H_{0}+\Omega L_{z})}{\hbar \Omega }\right)
L_{m}\left( \frac{2(H_{0}-\Omega L_{z})}{\hbar \Omega }\right) \frac{1}{e^{%
\frac{E_{n,m}-\mu }{k_{B}T}}+1}  %\tag{35}
\end{equation}%
\end{widetext}
Since, in the $T\rightarrow 0$ limit the Fermi distribution function becomes
the step function, with $\lambda =\mu (T=0)$ as the Fermi energy, that is,%
\begin{equation}
\frac{1}{\exp \left( \frac{E_{n,m}-\mu }{k_{B}T}\right) +1}\rightarrow
\Theta (\lambda -E_{n,m}),  %\tag{36}
\end{equation}%
the result (35) reduces to the correct zero temperature limit given in Eq.
(25). As can be seen in (35), the Fermi distribution function enters in
a simple way in the expression of the phase space distribution. This
suggests to examine a similar situation for the case of bosons. In this
case, the first-order density matrix in spatial coordinates\ at temperature $%
T$ is%
\begin{equation}
\rho ^{B}({\bm r}+\frac{\bm s}2,{\bm r}-\frac{\bm s}2,T)=\sum_{n=0}^{\infty } 
\sum_{m=0}^{\infty}
\frac{\phi _{n,m}({\bm r}+\frac{\bm s}2)\phi _{n,m}^{\ast }(%
{\bm r}-\frac{\bm s}2)}{\exp \left( \frac{%
E_{n,m}-\mu }{k_{B}T}\right) -1},  %\tag{37}
\end{equation}%
where we have included the Bose distribution function. Following the same
derivation as done for Fermions, one immediatly gets for the phase space
density of bosons at finite temperature, the result%
\begin{widetext}
\begin{equation}
\rho _{W}^{B}({\bm r},{\bm p},T)=4e^{-\frac{2H_{0}}{%
\hbar \Omega }}\sum_{n=0}^{\infty }\sum_{m=0}^{\infty
}(-1)^{n+m}L_{n}\left( \frac{2(H_{0}+\Omega L_{z})}{\hbar \Omega }\right)
L_{m}\left( \frac{2(H_{0}-\Omega L_{z})}{\hbar \Omega }\right) \frac{1}{e^{%
\frac{E_{n,m}-\mu }{k_{B}T}}-1}.  %\tag{38}
\end{equation}%
\end{widetext}
This last equation may constitute a useful starting point to study
thermodynamical properties in phase space of charged Bose gas, in particular
at low temperatues.

\section{The autocorrelation function}

The autocorrelation function, also called the reciprocal form factor \cite{[19]}, is known to provide information on the off-diagonal part of the
density matrix $\rho ({\bm r},{\bm r}')$ and is
defined in spatial coordinates as
\begin{equation}
B({\bm s})=\int_{\mathbb R^2} \exp \left( -\frac{i{\bm p}\cdot {\bm s}}{\hbar }\right) n({\bm p})\, d{\bm p}  %\tag{39}
\end{equation}%
with ${\bm s}={\bm r}-{\bm r}'$ and $n(%
{\bm p})$ is the density profile in momentum space. The latter is
defined by a similar relation as in Eq. (30), where one has to convert the
normalized spatial wavefunctions $\phi _{n,m}$ into their analogues $%
\widetilde{\phi }_{n,m}({\bm p})$ in momentum space, that is%
\begin{equation}
n({\bm p})=\sum_{n=0}^{\infty }\sum_{m=0}^{\infty }%
\widetilde{\phi }_{n,m}({\bm p})\widetilde{\phi }_{n,m}^{\ast }(%
{\bm p})\Theta (\lambda -E_{n,m}).  %\tag{40}
\end{equation}%
On the other hand the momentum density $n({\bm p})$ can also be
obtained through the Wigner phase space distribution%
\begin{equation}
n({\bm p})=\int_{\mathbb R^2} \rho _{W}({\bm r},{\bm p})%
\frac{d{\bm r}}{(2\pi \hbar )^{2}}  %\tag{41}
\end{equation}%
and is normalized to the total particle number $N$ of the system%
\begin{equation}
\int_{\mathbb R^2} n({\bm p})\ d{\bm p}=N.  %\tag{42.a}
\end{equation}%
Therefore, it follows from Eq. (39), that 
$B({\bm 0})=N$. In the following we shall derive a closed analytical result for $B(%
{\bm s})$. To do so, we first insert Eq. (23) into Eq. (41), to
obtain%
\begin{equation}
n({\bm p})=\frac{1}{2\pi i}\int_{c-i\infty }^{c+i\infty
}d\beta \frac{e^{\beta \lambda }}{\beta }\int_{\mathbb R^2} \frac{d{\bm r}}{%
(2\pi \hbar )^{2}}C_{W}({\bm r},{\bm p},\beta ) 
%\tag{43}
\end{equation}%
to carry out the ${\bm r}$ integration, we use the analytical
form of $C_{W}({\bm r},{\bm p},\beta )$ given in Eq.
(10) and we rewrite it as follows%
\begin{eqnarray}
C_{W}({\bm r},{\bm p},\beta )&=&\frac{1}{\cosh \frac{%
\beta \hbar \Omega _{+}}{2}\cosh \frac{\beta \hbar \Omega _{-}}{2}}\exp
\left[ -\frac{fg}{f+gu^{2}}{\bm p}^{2}\right] \\
&&\times \exp \left[
-\left( f+gu^{2}\right) \left( {\bm r}+\frac{gu}{f+gu^{2}}(%
{\bm p}\times {\bm k})\right) ^{2}\right]  \nonumber %\tag{44}
\end{eqnarray}%
where we have used $({\bm p}\times {\bm k})^{2}=%
{\bm p}^{2}$, since ${\bm p}$ is a planar vector. The
above result can now be inserted into Eq. (43), to obtain 
\begin{widetext}
\begin{eqnarray}
n({\bm p})&=&\frac{1}{2\pi i}\int_{c-i\infty }^{c+i\infty
}d\beta \frac{\exp (\beta \lambda )}{(2\pi \hbar )^{2}\beta }\left[ \frac{1}{%
\cosh \frac{\beta \hbar \Omega _{+}}{2}\cosh \frac{\beta \hbar \Omega _{-}}{2%
}}\frac{\pi }{\left( f+gu^{2}\right) }\exp \left( -\frac{fg}{f+gu^{2}}%
{\bm p}^{2}\right) \right] \nonumber \\ %\tag{45}
&=&\frac{1}{2\pi i}\int_{c-i\infty }^{c+i\infty
}d\beta\ \frac{\exp (\beta \lambda )}{(2\pi \hbar )^{2}\beta }\left[ \frac{%
2\pi \hbar }{m^{\ast }\Omega \sinh (\beta \hbar \Omega )}\exp \left( -\frac{%
2\sinh \frac{\beta \hbar \Omega _{+}}{2}\sinh \frac{\beta \hbar \Omega _{-}}{%
2}}{\hbar m^{\ast }\Omega \sinh (\beta \hbar \Omega )}{\bm p}%
^{2}\right) \right], %\tag{46}
\end{eqnarray}%
where we have used (11). Putting this result with this present form into Eq. (39) and using Eq. (9), one then finds%
\begin{eqnarray}
B({\bm s})&=&\frac{1}{2\pi i}\int_{c-i\infty }^{c+i\infty
}d\beta \frac{\exp (\beta \lambda )}{2\pi \hbar m^{\ast }\Omega \beta \sinh
(\beta \hbar \Omega )}\left[ \int_{\mathbb R^2} d {\bm p}\exp \left( -%
\frac{2\sinh \frac{\beta \hbar \Omega _{+}}{2}\sinh \frac{\beta \hbar \Omega
_{-}}{2}}{\hbar m^{\ast }\Omega \sinh (\beta \hbar \Omega )}{\bm p%
}^{2}-\frac{i{\bm p}\cdot {\bm s}}{\hbar }\right) \right] \nonumber \\
%\tag{47}
&=&\frac{1}{2\pi i}\int_{c-i\infty }^{c+i\infty
}d\beta \frac{\exp (\beta \lambda )}{\beta }\left[ \frac{1}{4\sinh \frac{%
\beta \hbar \Omega _{-}}{2}\sinh \frac{\beta \hbar \Omega _{+}}{2}}\exp
\left( -\frac{m^{\ast }\Omega }{8\hbar }\frac{\sinh (\beta \hbar \Omega )}{%
\sinh \frac{\beta \hbar \Omega _{+}}{2}\sinh \frac{\beta \hbar \Omega _{-}}{2%
}}{\bm s}^{2}\right) \right] .  %\tag{48}
\end{eqnarray}%
\end{widetext}
Remember that, $\Omega =(\Omega _{+}+\Omega _{-})/2$, so that%
\begin{equation}
\frac{\sinh (\beta \hbar \Omega )}{\sinh \frac{\beta \hbar \Omega _{+}}{2}%
\sinh \frac{\beta \hbar \Omega _{-}}{2}}=\coth \frac{\beta \hbar \Omega _{+}%
}{2}+\coth \frac{\beta \hbar \Omega _{-}}{2}  %\tag{49}
\end{equation}%
plugging this result into the exponential of Eq. (46), to get%
\begin{eqnarray}
B({\bm s})&=&\frac{1}{2\pi i}\int_{c-i\infty }^{c+i\infty
}d\beta \frac{e^{\beta \lambda}}{\beta }\left[ \frac{\exp \left( -\frac{%
m^{\ast }\Omega }{8\hbar }\left( \coth \frac{\beta \hbar \Omega +}{2}\right) 
{\bm s}^{2}\right) }{2\sinh \frac{\beta \hbar \Omega _{+}}{2}}\right. \nonumber \\
&& \qquad \qquad \left.
\times \frac{\exp \left( -\frac{m^{\ast }\Omega }{8\hbar }\coth \left( \frac{%
\beta \hbar \Omega _{-}}{2}\right) {\bm s}^{2}\right) }{2\sinh 
\frac{\beta \hbar \Omega _{-}}{2}}\right]  %\tag{50}
\end{eqnarray}%
At this level, we can carry out explicitly the inverse Laplace transform by
first using the following expansion in terms of Laguerre polynomials \cite{[16]}%
\begin{equation}
\frac{\exp \left[ -x\coth \left( \frac{\beta \hbar \Omega _{\pm }}{2}\right) %
\right] }{\sinh (\frac{\beta \hbar \Omega _{\pm }}{2})}=2e^{-x}\sum_{n=0}^{\infty }L_{n}(2x)\ e^{ -\beta \hbar \Omega _{\pm }(n+\frac{1}{2})}  %\tag{51}
\end{equation}%
and followed by Eq. (24), one then finds%
\begin{equation}
B({\bm s})=e^{-\frac{m^{\ast }\Omega }{4\hbar }{\bm s}%
^{2}}\sum_{m,n=0}^{\infty } L_{n}\left( \frac{%
m^{\ast }\Omega }{4\hbar }{\bm s}^{2}\right) L_{m}\left( \frac{%
m^{\ast }\Omega }{4\hbar }{\bm s}^{2}\right) \Theta (\lambda
-E_{n,m})  %\tag{52}
\end{equation}%
Notice that, an interesting feature of the above autocorrelation function is
that it is expressed in terms of Laguerre polynomials with same arguments,
note also that it is isotropic in spatial coordinates, i.e, depends only on
the length $\left\vert {\bm s}\right\vert $of the vector $%
{\bm s}$. Setting ${\bm s}=0$, one obtains $B({\bm 0})=N$, as it is required. 
In Fig.~2 we display the $T=0$ spatial dependence of the autocorrelation function for $N=20$ particles, choosing the following values of the parameters $\omega_0/\omega_L=1$ and Fermi energy $\lambda=6.35\hbar\omega_L$.

\begin{figure}[ht]%
\centering
\includegraphics[width=5cm]{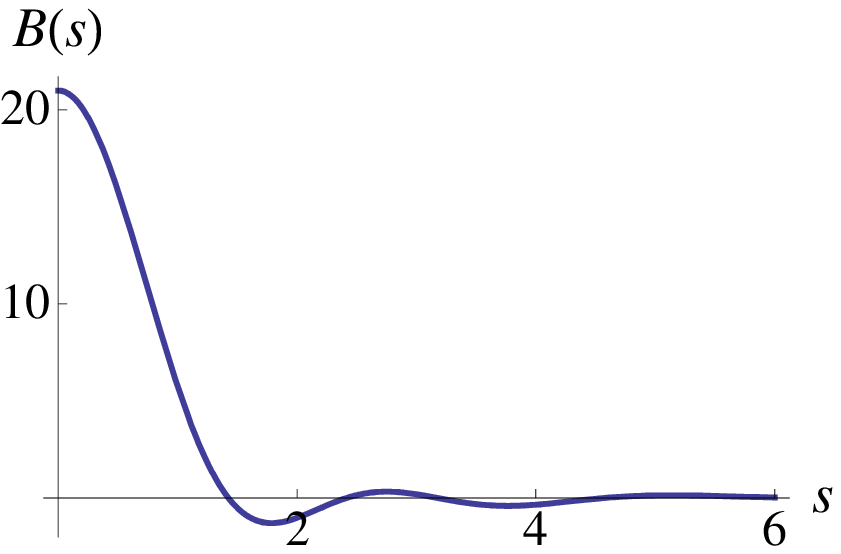} 
\caption{The $T=0$ autocorrelation function $B({\bm s})=B(s)$ for $N=20$ particles, with $\omega_0/\omega_L=1$ and Fermi energy $\lambda=6.35\hbar\omega_L$. Lengths are plotted in units of the magnetic length $l=\sqrt{\hbar c/eB}$.}
\end{figure}

Finally, let us come to
the finite temperature expression of $B({\bm s})$. In a similar
way as was done in the previous section, one can immediatly write down its
expression. For Fermions, this is simply achieved by replacing the $\Theta
(\lambda -E_{n,m})$ in Eq. (50) by the Fermi function%
\begin{equation}
B({\bm s})=e^{-\frac{m^{\ast }\Omega }{4\hbar }{\bm s}%
^{2}}\sum_{m,n=0}^{\infty }  \frac{L_{n}\left( \frac{%
m^{\ast }\Omega }{4\hbar }{\bm s}^{2}\right) L_{m}\left( \frac{%
m^{\ast }\Omega }{4\hbar }{\bm s}^{2}\right)}{\exp \left( 
\frac{E_{n,m}-\mu }{k_{B}T}\right) +1}.  %\tag{53}
\end{equation}%
For Bosons, all that is required is the replacement of the Fermi function by
the Bose function
\begin{equation}
B({\bm s})=e^{-\frac{m^{\ast }\Omega }{4\hbar }{\bm s}%
^{2}}\sum_{m,n=0}^{\infty }
\frac{L_{n}\left( \frac{%
m^{\ast }\Omega }{4\hbar }{\bm s}^{2}\right) L_{m}\left( \frac{%
m^{\ast }\Omega }{4\hbar }{\bm s}^{2}\right)}{\exp \left( 
\frac{E_{n,m}-\mu }{k_{B}T}\right) -1}.  %\tag{54}
\end{equation}

\section{Strong magnetic field case}

The strong magnetic field case at low temperature is of particular interest
in quantum dots. In this limit only a few Landau levels are occupied and the
magnetic field length $l=(\hbar c/eB)^{1/2}$ is small leading to slowly
varying confining external harmonic potential on the scale of $l$. In what
follows we shall examine the strong magnetic field (SB) case, where one has $%
\omega _{0}/\omega _{L}\ll 1$ then $\Omega \approx \omega _{L}$, $\Omega
_{+}\approx 2\omega _{L}$ and $\Omega _{-}\approx $ $\omega
_{0}^{2}/(2\omega _{L})$ which yields respectively for the functions given
in Eq. (11), to the leading order 
\begin{equation}
f(\beta)\approx \beta m^{\ast }\omega _{0}^{2}/2,\ g(\beta)\approx \frac{\tanh
(\beta \hbar \omega _{L})}{2m^{\ast }\hbar \omega _{L}},\ u(\beta)\approx
m^{\ast }\omega _{L} . %\tag{55}
\end{equation}%
Substituting this results into Eq. (10), gives immediately the result for
the Wigner transform of the Bloch density
\begin{equation}
C_{W}^{SB}({\bm r},{\bm p},\beta )=
\frac{e^{-\beta m^{\ast }\omega _0^2{\bm r}^2/2}}{\cosh \beta \hbar \omega _{L}}
\exp \left[ -\frac{\tanh \beta \hbar
\omega _{L}}{2 m^{\ast }\hbar \omega _{L}}\left( {\bm p}+%
\frac{e}{c}{\bm A}\right) ^{2}\right]  . %\tag{56}
\end{equation}%
In this case, the net result we get is a product of the Bloch density of
free charged particles in magnetic field  [see Eq. (13)]
and an exponential factor limiting the spatial distribution. Let us now,
calculate the corresponding Wigner phase space density. Inserting Eq. (54)
into Eq. (23) and making use of Eqs. (19) and (24) , one obtains%
\begin{eqnarray}
\rho _{W}^{SB}({\bm r},{\bm p})&=& 2\exp \left[ -\frac{%
H_{magn}}{\hbar \omega _{L}}\right] \sum_{n=0}^{\infty
}(-1)^{n}L_{n}\left( \frac{2H_{magn}}{\hbar \omega _{L}}\right) \\
&& \qquad \qquad\times \Theta
\left( \lambda -(2n+1)\hbar \omega _{L}-\frac{m^{\ast }\omega _{0}^{2}}{2}%
{\bm r}^{2}\right) , \nonumber %\tag{57}
\end{eqnarray}%
where $H_{magn}=({\bm p}+(e/c){\bm A})^{2}/(2m)$
is the Hamiltonian for a particle in the presence of the magnetic field
alone. One immediately recognizes in the argument of the Heaviside function
the discrete Landau level energies $(2n+1)\hbar \omega _{L}$. As can be
seen, the phase space density in above has a simple analytical form,
therefore we can easily obtain, in this high magnetic field limit, the
corresponding density matrix $\rho ({\bm r}+{\bm s}/2,%
{\bm r}-{\bm s}/2)$ in spatial coordinates. To do so,
we make use of the inverse Wigner transformation. According to Eq. (6), one
has 
\begin{widetext}
\begin{equation}
\rho ^{SB}({\bm r}+{\bm s}/2,{\bm r}-%
{\bm s}/2) = 2\sum_{n=0}^{\infty }(-1)^{n}\Theta \left(
\lambda -(2n+1)\hbar \omega _{L}-\frac{m^{\ast }\omega _{0}^{2}}{2}%
{\bm r}^{2}\right) \int_{\mathbb R^2} \frac{d {\bm p}}{(2\pi \hbar )^{2}}e^{+i%
{\bm p} \cdot {\bm s}/\hbar }e^{-\frac{H_{magn}}{\hbar
\omega _{L}}}L_{n}\left( \frac{2H_{magn}}{\hbar \omega _{L}}\right) .  %\tag{58}
\end{equation}%
\end{widetext}
The last integral can be carried out as follows. Denoting it by $I$ and
using the canonical momentum ${\bm K}={\bm p}+(e/c)%
{\bm A}$, one obtains%
\begin{equation}
I=e^{-i\frac{e}{\hbar c}{\bm A}\cdot {\bm s}}\int_{\mathbb R^2} \frac{d {\bm K}}{(2\pi \hbar )^{2}}
e^{i{\bm K} \cdot {\bm s}/\hbar }e^{-\frac{{\bm K}^{2}}{2m^{\ast }\hbar
\omega _{L}}}L_{n}\left( \frac{{\bm K}^{2}}{m^{\ast }\hbar \omega
_{L}}\right) , %\tag{59}
\end{equation}%
and changing to the variable $t=K/\sqrt{m^{\ast }\hbar \omega _{L}}$, we get 
\begin{eqnarray}
I &=&\frac{m^{\ast }\hbar \omega _{L}}{(2\pi \hbar )^{2}}e^{-i\frac{e}{\hbar
c}{\bm A}\cdot{\bm s}}\int_{0}^{\infty }te^{-\frac{t^{2}}{%
2}}L_{n}\left( t^{2}\right) dt\int_{0}^{2\pi }d\phi
e^{i\sqrt{\frac{m^{\ast}\omega _{L}}\hbar} ts\cos \phi}  \nonumber \\
&=&\frac{e^{-i\frac{e}{\hbar c}{\bm A}.{\bm s}}}{4\pi
l^{2}}\int_{0}^{\infty }te^{-\frac{t^{2}}{2}}L_{n}\left( t^{2}\right) J_{0}(%
\sqrt{m^{\ast }\omega _{L}/\hbar }\ ts)dt , %\tag{60}
\end{eqnarray}
where we have made use of the relation \cite{[16]} 
\begin{equation}
\int_{0}^{2\pi }d\phi e^{ix\cos (\phi )}=2\pi J_{0}(x) \nonumber %\tag{61}
\end{equation}%
to get the last line, $J_{0}(x)$ being the Bessel function. The following
relation \cite{[16]}%
\begin{equation}
\int_{0}^{\infty }xe^{-\frac{x^{2}}{2}}L_{n}\left( x^{2}\right)
J_{0}(xy)dx=(-1)^{n}e^{-\frac{y^{2}}{2}}L_{n}\left( y^{2}\right)  \nonumber%\tag{62}
\end{equation}%
helps to perform the integral in Eq. (58), to find%
\begin{equation}
I=\frac{(-1)^{n}}{2\pi l^{2}}e^{-i\frac{e}{\hbar c}{\bm A}.%
{\bm s}}e^{-\frac{m^{\ast }\omega _{L}}{2\hbar }{\bm s}%
^{2}}L_{n}\left( \frac{m^{\ast }\omega _{L}}{\hbar }{\bm s}%
^{2}\right)  %\tag{63}
\end{equation}%
Substituting this result into Eq. (56), yields%
\begin{eqnarray}
\rho ^{SB}({\bm r}+\frac{\bm s}2,{\bm r}-\frac{\bm s}2)&=&\frac{e^{-i\frac{e}{\hbar c}{\bm A}\cdot {\bm s}}}{2\pi l^{2}}e^{-\frac{m^{\ast }\omega _{L}}{2\hbar }%
{\bm s}^{2}}\sum_{n=0}^{\infty }L_{n}\left( \frac{m^{\ast
}\omega _{L}}{\hbar }{\bm s}^{2}\right) \nonumber \\
&& \times \Theta \left( \lambda
-(2n+1)\hbar \omega _{L}-m^{\ast }\omega _{0}^{2}{\bm r}%
^{2}/2\right) . \nonumber %\tag{64}
\end{eqnarray}%
The local density is obtained by setting ${\bm s}={\bm %
0}$, 
\begin{equation}
\rho ^{SB}({\bm r})=\frac{1}{2\pi l^{2}}\sum_{n=0}^{\infty
}\Theta \left( \lambda -(2n+1)\hbar \omega _{L}-\frac{m^{\ast }\omega
_{0}^{2}}{2}{\bm r}^{2}\right)  .%\tag{65}
\end{equation}%
In their study of a two dimensional electron gas subjected to a magnetic field and partially 
confined by a harmonic potential, the authors of Ref.~\cite{[20]} obtained a similar
result for the spatial density using a different approach (remark that their parabolic
potential is taken only in the $x$ direction, i.e, $V(x,y)=m^{\ast }\omega
_{0}^{2}x^{2}/2$). As noticed by these authors, the density contains
compressible and incompressible regions.

In order to rewrite Eq. (60) in a more compact form, we use the following identity relating the Heaviside and the integer part functions
\begin{equation}
\sum_{n=0}^{\infty }\Theta (x-n)=\Theta (x+1)\text{Int}(x+1).  %\tag{66}
\end{equation}%
Then, $\rho ({\bm r}) $ becomes 
\begin{eqnarray}
\rho ({\bm r})&=&\frac{1}{2\pi l^{2}}\Theta \left( \frac{\lambda
-m^{\ast }\omega _{0}^{2}{\bm r}^{2}/2+\hbar \omega _{L}}{2\hbar
\omega _{L}}\right) \nonumber \\
&& \qquad \times\text{Int}\left( \frac{\lambda -m^{\ast }\omega _{0}^{2}%
{\bm r}^{2}/2+\hbar \omega _{L}}{2\hbar \omega _{L}}\right) .
%\tag{67}
\end{eqnarray}%
In Fig.~3 we plot the above zero-temperature spatial density for the case of $N=200$ particles with parameters $\omega_0/\omega_L=0.2048$ and Fermi energy $\lambda= 4.1\hbar\omega_L$

\begin{figure}[h]%
\centering
\includegraphics[width=5cm]{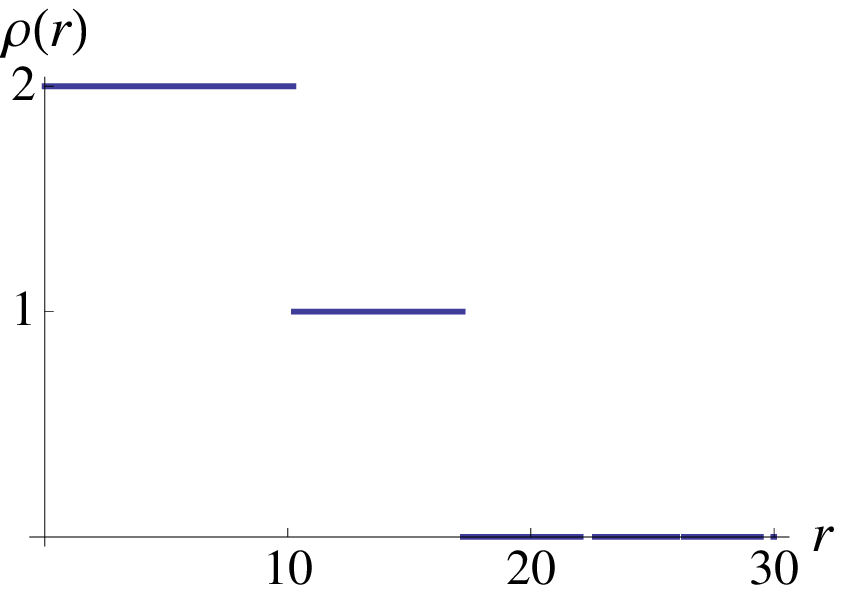} 
\caption{Plot of the spatial density $\rho ({\bm r})=\rho ( r)$ in units of $(2\pi l^{2})^{-1}$ at $T=0$ in a harmonic oscillator potential with $\omega_0/\omega_L=0.2048$ and Fermi energy $\lambda= 4.1\hbar\omega_L$, corresponding to $N=200$ particles. Lengths are plotted in units of the magnetic length $l=\sqrt{\hbar c/eB}$.}
\end{figure}

For ultra strong magnetic field, such that all the particles reside in the
lowest Landau level (LLL), Eq. (60) reduces to%
\begin{equation}
\rho ^{LLL}({\bm r})=\frac{1}{2\pi l^{2}}\Theta \left( \lambda
-\hbar \omega _{L}-m^{\ast }\omega _{0}^{2}{\bm r}^{2}/2\right) 
%\tag{68}
\end{equation}

Before closing this section, it is intersting to calculate the momentum
density or the momentum distribution $n({\bm p})$ in the
strong magnetic field case. This important distribution was already
introduced in the previous section but its calculation has not been fully
completed since this density was used there as an intermediate to obtain the
autocorrelation function. In a short, we start from its expression given in
Eq. (46) and introduce the strong magnetic field approximations we used
above, namely $\sinh \left( \beta \hbar \Omega _{+}/2\right) \approx \sinh
\beta \hbar \omega _{L}$, $\sinh \left( \beta \hbar \Omega _{-}/2\right)
\approx $ $\beta \hbar \omega _{0}^{2}/(4\omega _{L})$ and $\sinh \left(
\beta \hbar \Omega \right) \approx \sinh \beta \hbar \omega _{L}$ to get%
\begin{equation}
n^{SB}({\bm p})=\frac{1}{2\pi i}\int_{c-i\infty
}^{c+i\infty }d\beta \frac{\exp (\beta \lambda )}{(2\pi \hbar )m^{\ast
}\omega _{L}\beta }
\frac{\exp
\left( -\beta \frac{\omega _{0}^{2}}{2m^{\ast }\omega _{L}^{2}}%
{\bm p}^{2}\right)}{\sinh \beta \hbar \omega _{L}} . %\tag{69}
\end{equation}%
At this level, it is easy to perform the inverse Laplace transform by using
the expansion 
\begin{equation}
\frac{1}{\sinh (\beta \hbar \omega _{L})}=2\sum_{n=0}^{\infty }\exp %
\left[ -(2n+1)\beta \hbar \omega _{L}\right]  \nonumber %\tag{70}
\end{equation}%
with Eq. (24), which results in 
\begin{equation}
n^{SB}({\bm p})=\frac{2l^{2}}{\pi \hbar ^{2}}\sum_{n=0}^{%
\infty }\Theta \left( \lambda -(2n+1)\hbar \omega _{L}-\frac{\omega _{0}^{2}%
}{2m^{\ast }\omega _{L}^{2}}{\bm p}^{2}\right) . %\tag{71}
\end{equation}%
Like the spatial density in Eq. (60), the momentum density exhibites the
same structure consisting of a series of wide steps in momentum space.

The above results have been obtained for spinless charged particles at $T=0$
and the generalization to finite temperatures can be done without any
particular difficulties.

\section{Summary and outlook}

We have derived some simple exact closed expressions for the Wigner
transform of the canonical Bloch density of two-dimensional harmonic
oscillator in a uniform magnetic field. We have also obtained exact
analytical form for the Wigner phase space density at zero and nonzero
temperature. Our results are valid for arbitrary magnetic field strengths
and hold for both Fermions and Bosons. For the system under study, we have
found simple and exact analytical expression for the so called
autocorelation function. The high magnetic field case has been examined. Our
investigation in phase space complement the recent works in spatial
coordinates. The results we obtained would constitute useful starting point
for the study, in phase space, of thermodynamical properties in the field of
cold atom gases.

\begin{acknowledgments}
This work has been partially supported by the Spanish Ministerio de Educaci\'on y Ciencia  (Project MTM2005-09183) and Junta de Castilla y Le\'on (Excellence Project GR224). KB gratefully acknowledges the hospitality
and the financial support granted to him during his stay at the Departamento
de F\'{\i}sica Te\'orica, At\'omica y \'Optica, Valladolid (Spain), where part of this work was done.
\end{acknowledgments}

\appendix

\section{}

The purpose of this appendix is to derive the expression of the function $%
G\left( {\bm r},{\bm p},\beta \right) $ given in Eq. (17). First, we evaluate the various functions of $\beta $ in Eq. (15), namely $U(\beta )=\left( f(\beta )+g(\beta )u^{2}(\beta )\right) $, $g(\beta)$ and $V(\beta )=2g(\beta )u(\beta )$. Using Eq. (11), we get for $U(\beta) $ 
\begin{widetext}
\begin{equation}
U(\beta ) =\frac{2m^{\ast }\Omega \sinh   \frac{\beta \hbar \Omega
_{-}}{2}  \sinh  \frac{\beta \hbar \Omega _{+}}{2}  }{%
\hbar \sinh (\beta \hbar \Omega )}+\frac{m^{\ast }\Omega \sinh ^{2} 
\beta \hbar \omega _{L}}{2\hbar \sinh (\beta \hbar \Omega )\cosh
 \frac{\beta \hbar \Omega _{-}}{2} \cosh  \frac{\beta
\hbar \Omega _{+}}{2}} 
=\frac{m^{\ast }\Omega/2\hbar}{\sinh \beta \hbar \Omega}  \frac{%
\sinh \left( \beta \hbar \Omega _{-}\right) \sinh \left( \beta \hbar \Omega
_{+}\right) +\sinh ^{2} \beta \hbar \omega _{L}}{\cosh  
\frac{\beta \hbar \Omega _{-}}{2}  \cosh  \frac{\beta \hbar
\Omega _{+}}{2}} .
\end{equation}%
\end{widetext}
From $\sinh \left( \beta \hbar \Omega _{-}\right) \sinh \left( \beta
\hbar \Omega _{+}\right) =\frac{1}{2}\left( \cosh (\beta \hbar (\Omega
_{-}+\Omega _{+})\right) -\cosh (\beta \hbar (\Omega _{-}-\Omega _{+}))$, and $\Omega _{-}+\Omega _{+}=$ $2\Omega $ , $\Omega _{-}-\Omega
_{+}=-2\omega _{L}$, we obtain%
\begin{eqnarray*}
U(\beta ) &=&\frac{m^{\ast }\Omega \left[ \cosh 2\beta \hbar \Omega -\cosh
\beta \hbar \omega _{c}+2\sinh ^{2} \beta \hbar \omega _{L} \right] }{4\hbar \sinh (\beta \hbar \Omega )\cosh \left( \frac{\beta \hbar
\Omega _{-}}{2}\right) \cosh  \frac{\beta \hbar \Omega _{+}}{2}
} \\
&=&\frac{m^{\ast }\Omega \sinh \beta \hbar \Omega}{2\hbar \cosh 
\frac{\beta \hbar \Omega _{-}}{2} \cosh \frac{\beta \hbar
\Omega _{+}}{2}} .
\end{eqnarray*}%
If we use again $\Omega _{-}+\Omega _{+}=$ $2\Omega $, the above result
becomes 
\begin{equation}
U(\beta )=\frac{m^{\ast }\Omega }{2\hbar }\left( \tanh \frac{\beta \hbar
\Omega _{+}}{2}+\tanh \frac{\beta \hbar \Omega _{-}}{2}\right)  %\tag{A.1}
\end{equation}%
We now turn to the two remaining functions. For convenience, we rewrite the
function $g(\beta )$, given in Eq. (11), as%
\begin{equation}
g(\beta )=\frac{1}{2m^{\ast }\hbar \Omega }\left( \tanh \frac{\beta \hbar
\Omega _{+}}{2}+\tanh \frac{\beta \hbar \Omega _{-}}{2}\right) . %\tag{A.2}
\end{equation}%
For the function $V(\beta )$, one simply gets
\begin{equation*}
V(\beta )=\frac{\sinh 2\beta \hbar \omega _{L}}{\hbar \cosh \frac{\beta
\hbar \Omega _{-}}{2}\cosh \frac{\beta \hbar \Omega _{+}}{2}},
\end{equation*}%
and using $\ \Omega _{+}-\Omega _{-}=$ $2\omega _{L}$, one ends with%
\begin{equation}
V(\beta )=\frac{1}{\hbar }\left( \tanh \frac{\beta \hbar \Omega _{+}}{2}%
-\tanh \frac{\beta \hbar \Omega _{-}}{2}\right) . %\tag{A.3}
\end{equation}%
Substituting Eqs. (A1)--(A3) into Eq. (16), simple manipulations
yield to the desired result (17).

\section{}

In this appendix we shall show that, in the absence of magnetic field, the
phase space density in Eq. (28) reduce to the result in Eq. (29). In this
limit Eq. (28) becomes%
\begin{widetext}
\begin{equation}
\rho _{W}^{B=0}({\bm r},{\bm p})=4e^{-\frac{2H_{0}}{%
\hbar \omega _{0}}}\sum_{n=0}^{N_{+}}\sum_{m=0}^{N_{-}}(-1)^{n+m}L_{n}\left( \frac{2(H_{0}+\omega _{0}L_{z})}{%
\hbar \omega _{0}}\right) L_{m}\left( \frac{2(H_{0}-\omega _{0}L_{z})}{\hbar
\omega _{0}}\right) . %\tag{B.1}
\end{equation}%
Here, $N_{+}=\text{Int}(\frac{\lambda }{\hbar \omega _{0}})-1$ and $N_{-}=\text{Int}\left( 
\frac{\lambda }{\hbar \omega _{0}}-n-1\right) =N_{+}-n$. The physical
meaning of $N_{+}$ is only but the quantum number of the last occupied
harmonic oscillator shell, denoted by $M$ in Eq. (29). The Hamiltonian 
$H_{0}$ appearing in Eq. (B1) refers to Eq. (18) but with  $\Omega =\omega
_{0}$. Putting $p=$ $n+m$, Eq. (B1) rewrites%
\begin{equation}
\rho _{W}^{B=0}({\bm r},{\bm p})=4e^{-\frac{2H_{0}}{%
\hbar \omega _{0}}}\sum_{p=0}^{M}(-1)^{p}\sum_{m=0}^{p}L_{p-m}%
\left( \frac{2(H_{0}+\omega _{0}L_{z})}{\hbar \omega _{0}}\right)
L_{m}\left( \frac{2(H_{0}-\omega _{0}L_{z})}{\hbar \omega _{0}}\right)  ,
%\tag{B.2}
\end{equation}%
and using the identity 
$\sum_{m=0}^{p}L_{p-m}\left( x\right) L_{m}\left( y\right) =L_{p}^{1}(x+y)  %\tag{B.3}
$ (see \cite{[16]}), the result in Eq. (29) is then recovered.
\end{widetext}


\begin{thebibliography}{99}
\bibitem{[1]} 
L. Jacak, P. Hawrylak, and A. W\`ojs, Quantum Dots (Springer, Berlin, 1998).

\bibitem{[2]} 
T. Chakraborty, Quantum Dots: A Survey of the Properties of Artificial Atoms  (Elsevier, Amsterdam, 1999).

\bibitem{[3]} 
B. De Marco and D. S. Jin, Science \textbf{285} 1703
(1999);\ A. G. Truscott, K. E. Strecker, W. I. McAlexander, G. B. Partridge,
and R. G. Hulet, Science \textbf{291} 2570 (2001); F. Schreck et al, Phys.
Rev. Lett.\textbf{87} 080403 (2001); S. R. Granade, M.E. Gehm, K.M. OHara, and J.E. Thomas, Phys. Rev. Lett. \textbf{88} 120405 (2002); G. Roati, F. Riboli, G. Modugno, M. Inguscio, Phys. Rev. Lett. \textbf{89}
150403 (2002); Z. Hadzibabic et al, Phys. Rev. Lett. \textbf{91 }160401
(2003)

\bibitem{[4]} 
P. Vignolo, A. Minguzzi and M.P. Tosi, Phys. Rev. Lett. \textbf{85} 2850 (2000); F. Gleisberg, W. Wonneberger, U. Schl\"{o}der and C. Zimmermann, Phys. Rev. A \textbf{62} 063602 (2000); N.H. March and L.M. Nieto, Phys. Rev. A \textbf{63} 044502 (2001); N.H. March, L.M. Nieto and M.P. Tosi, Physica B \textbf{293} 308 (2001); X.Z. Wang, Phys. Rev. A \textbf{65} 045601 (2002); E. J. Mueller, Phys. Rev. Lett. \textbf{93} 190404 (2004); A. Minguzzi, S. Succi, F. Toschi, M.P. Tosi and P. Vignolo, Phys. Rep. \textbf{395} 223 (2004).

\bibitem{[5]} 
M. Brack and B.P. van Zyl, Phys. Rev. Lett \textbf{86} 1574
(2001); M. Brack and M. V. N. Murthy, J. Phys. A: Math. Gen. \textbf{36}
1111 (2003); B.P. van Zyl, R. K. Bhaduri, A. Suzuki and M. Brack, Phys. Rev. A 
\textbf{67} 023609 (2003); B.P. van Zyl and D. A. W. Hutchinson, Phys. Rev. B 
\textbf{69} 024520 (2004).

\bibitem{[6]} 
P. Shea and B.P. van Zyl, Phys. Rev. B \textbf{74} 205334
(2006); P. Shea and B.P. van Zyl, J. Phys. A: Math. Theor. \textbf{40} 10589
(2007); P. Shea and B.P. van Zyl, J. Phys. A: Math. Theor. \textbf{41} 135305
(2008).

\bibitem{[7]} 
E.P. Wigner, Phys. Rev. \textbf{40 }749 (1932)

\bibitem{[8]} 
See, for instance R. G. Parr and W. Yang, Density Functional
Theory of Atoms and Molecules (Oxford Science, New York, 1989).

\bibitem{[9]} 
M. Brack, R.K. Bhaduri, Semiclassical Physics, Frontiers in
Physics, vol. 96, Westview, Boulder, 2003

\bibitem{[10]} 
D. Leibfried, D. M. Meekhof, B. E. King, C. Monroe, W. M.
Itano and D. J. Wineland, Phys. Rev. Lett. \textbf{77} 4281 (1996).

\bibitem{[11]} 
N.H. March and A.M. Murray, Phys. Rev. \textbf{120} 830 (1960).

\bibitem{[12]} 
N.H. March and M.P. Tosi, J. Phys. A: Math. Gen. \textbf{18} L643
(1985).

\bibitem{[13]} 
W. Kohn, Phys. Rev. \textbf{123} 1242 (1961).

\bibitem{[14]} 
See, for instance, M. Hillery, R.F. O'Connel, M.O. Scully and E.P. Wigner, Phys. Rep. \textbf{106} 121 (1984).

\bibitem{[15]} 
E. H. Sondheimer and A. H. Wilson, Proc. R. Soc. London A 
\textbf{210} 173 (1951).

\bibitem{[16]} 
I. S. Gradshteyn and I. M. Ryzhik, Table of Integrals,
Series, and Products (Academic Press, New York, 1994, 5th edition).

\bibitem{[17]} 
M. Abramowitz and I. A. Stegun, Handbook of Mathematical
Functions (Dover Publications, New York, 1970, 9th printing)

\bibitem{[18]} 
S. Shlomo and M. Prakash, Nucl. Phys. A \textbf{357} 157
(1981).

\bibitem{[19]} 
See, for instance, P. Krusius, H. Is\"{o}mki and B. Kramer,
Phys. Rev. B \textbf{19} 1818 (1979).

\bibitem{[20]} 
M. R. Geller and G. Vignale, Phys. Rev. B \textbf{50} 11714
(1994).

\end{thebibliography}
\end{document}